\documentclass[reprint,amsmath, superscriptaddress, amssymb, aps, prm]{revtex4-1}
\usepackage{lipsum}  
\usepackage{stackengine}
\usepackage{graphicx}
\usepackage{hyperref}
\usepackage{amsmath}
\usepackage{soul}
\usepackage{dcolumn}
\usepackage{lipsum}  
\usepackage{float}
\usepackage{comment}
\usepackage{bm}
\usepackage{color}
\usepackage[normalem]{ulem}
\usepackage{lineno}
\begin{document}
\renewcommand{\thesection}{\Roman{section}}
\preprint{APS/123-QED}
\title{Stability of Wigner crystals and Mott insulators in twisted moiré structures}
\author{Daniel Erkensten}
\affiliation{Department of Physics, Philipps-Universit{\"a}t Marburg, 35037 Marburg, Germany}
\author{Samuel Brem}
\affiliation{Department of Physics, Philipps-Universit{\"a}t Marburg, 35037 Marburg, Germany}
\author{Raul Perea-Causin}
\affiliation{Department of Physics, Stockholm University, 10691 Stockholm, Sweden}
\author{Ermin Malic}
\affiliation{Department of Physics, Philipps-Universit{\"a}t Marburg, 35037 Marburg, Germany}
\begin{abstract}
Transition metal dichalcogenides (TMDs) constitute an intriguing platform for studying charge-ordered states including conventional and generalized Wigner crystals as well as Mott insulating states. In this work, we combine a phonon mode expansion of the electronic crystal vibrations with the Lindemann criterion to investigate the quantum and thermal stability of these strongly correlated phases in the exemplary materials of MoSe$_2$ monolayers and twisted MoSe$_2$-WSe$_2$ heterostructures. We find that the moiré potential in heterobilayers acts as a harmonic trap, flattening the energy dispersion of phonon excitations and resulting in an order of magnitude larger melting temperatures compared to  monolayer Wigner crystals. Furthermore, we explore the tunability of the correlated states with respect to dielectric environment and bilayer stacking. In particular, we show that the reduced screening in free-standing TMDs results in a tenfold increase in the melting temperature compared to hBN-encapsulated TMDs. Moreover, the deeper moiré potential in R-type stacked heterostructures makes generalized Wigner crystals more stable than in H-type stacking. Overall, our study provides important microscopic insights on the stability and tunability of charge-ordered states in TMD-based structures. 
\end{abstract}
\maketitle
\section{Introduction}
\begin{figure}[t!]
    \centering
    \includegraphics[width=\linewidth]{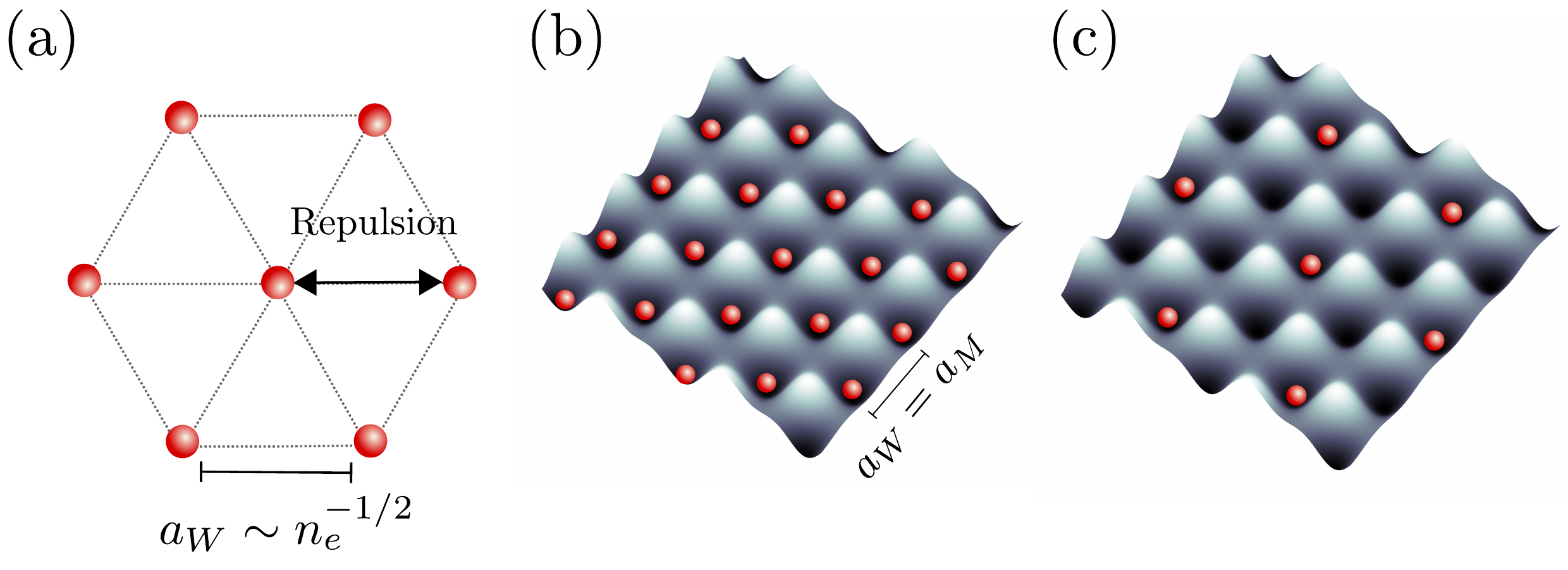}
    \caption{Charge-ordered states in atomically thin semiconductors. \textbf{(a)} Two-dimensional triangular Wigner crystal of electrons formed due to a strong Coulomb repulsion. The separation between the electrons, i.e. the Wigner lattice constant, $a_W$, is set by the electron density $n_e$. \textbf{(b)} Mott insulating state (filling factor $\nu=1$) formed in a moiré superlattice. Here, the separation between electrons is set by the moiré period $a_M$. \textbf{(c) } Generalized Wigner crystal state (filling factor $\nu=\frac{1}{3}$) formed in a moiré superlattice with the lattice constant $a_W=\sqrt{3} a_M$.}
    \label{schematicfig}
\end{figure}

Almost a century ago, Eugene Wigner predicted that an electron gas crystallizes at sufficiently low carrier densities and temperatures and builds the so-called Wigner crystal \cite{PhysRev.46.1002}. Concretely, the Coulomb repulsion between charges is so strong that it overcomes their kinetic energy, leading to the formation of a triangular lattice (Fig. \ref{schematicfig}(a)), where the lattice constant is set by the carrier density. The formation of two-dimensional Wigner crystals turns out to be particularly favorable in atomically thin semiconductors, such as transition-metal dichalcogenides (TMDs) \cite{smolenski2021signatures, zhou2021bilayer}, in which the weak dielectric screening results in an exceptionally strong Coulomb interaction \cite{perea2022exciton, mueller2018exciton, mak2022semiconductor, brem2022terahertz, wagner2021nonclassical, rosati2021dark}. Moreover, large effective masses in these materials lead to a small kinetic energy of electrons such that no external magnetic fields are necessary to realize a Wigner crystal \cite{smolenski2021signatures}. In addition, vertically stacking and/or twisting two or several TMD monolayers  gives rise to a periodic moiré potential \cite{mak2022semiconductor, brem2020tunable, tran2019evidence}. Here, carriers can be trapped in deep potential pockets resulting in an additional flattening of electronic bands. This facilitates the formation of other correlated phases of matter beyond regular Wigner crystals including Mott insulators (Fig. \ref{schematicfig}(b)) and generalized Wigner crystals appearing at certain fillings of the moiré lattice (Fig. \ref{schematicfig}(c)) \cite{regan2020mott, xu2020correlated, pan2020quantum, PhysRevB.103.125146}. 

Signatures of Wigner crystals have recently been established experimentally by means of optical spectroscopy in MoSe$_2$ monolayers \cite{smolenski2021signatures} and TMD bilayers \cite{xu2020correlated} and correlated phases have also been directly probed in STM measurements \cite{li2021imaging, li2023wigner}. Intriguingly, it has been demonstrated that the stability of the charge-ordered states in monolayers and bilayers differ dramatically both in terms of thermal and quantum (i.e. density-induced) melting. In particular, Wigner crystals in TMD monolayers have been shown to be stable up to $\sim 10$ K at densities of $n_e\sim 10^{11}$ $\mathrm{cm}^{-2}$ \cite{smolenski2021signatures}, whereas Mott insulators and generalized Wigner crystals get stabilized by the moir\'e potential and can survive beyond 100 K and $n_e>10^{12}$ $\mathrm{cm}^{-2}$ \cite{xu2020correlated}. To this date, the existence of Wigner crystals has been established only for selected TMD materials, and their stability has not been systematically investigated in terms of a realistic and material-specific theory. In particular, the complex interplay between the Coulomb interaction and the moiré potential as well as the impact of twist angle and stacking on the formation of correlated material phases has so far remained vague. 

In this work, we investigate thermal and quantum melting of Wigner crystal and Mott insulating states in TMD mono- and bilayers and shed light on the interplay between Coulomb repulsion and moiré trapping determining the stability of these correlated material phases. We treat the Coulomb interaction and the moiré potential on a microscopic and material-specific footing and compute the phonon excitations of the charge-ordered states. These excitations are interpreted as vibrations of the electronic lattice and are taken as input for the Lindemann criterion commonly used to qualitatively estimate the melting of correlated states \cite{bedanov1985modified, zhou2021bilayer}.

We find remarkably distinct phase diagrams for 2D electron Wigner crystals, Mott insulating states, and generalized Wigner crystals, which we explain through an intricate interplay between moiré pinning and Coulomb repulsion. This interplay is traced back to the dispersion of the phonon excitations of the Wigner crystal, which are strongly renormalized when accounting for the moiré potential. The latter results in harmonic traps of charge carriers with a zero-point energy that exceeds the thermal energy. Finally, we demonstrate a large tunability of the stability of charge-ordered states by varying the strength of the Coulomb interaction and the moiré potential depth through dielectric engineering and modification of stacking configuration, respectively.

\section{Vibrational excitations of charge-ordered states}
To investigate the stability of strongly correlated phases in atomically thin semiconductors, we first compute the vibrational excitations of electron crystals within the harmonic approximation \cite{bonsall1977some}. Here, we extend  previous approaches \cite{bonsall1977some, goldoni1996stability} taking into account the Coulomb repulsion between carriers by including also the moiré potential present in (twisted) TMD bilayers. We start from the Hamiltonian
\begin{equation}
H=\sum_{n}[-\frac{\hbar^2}{2 m}\nabla^2_{\mathbf{r}_n}+U(\mathbf{r}_n)+\frac{1}{2}\sum_{n'\neq n}V(\mathbf{r}_n-\mathbf{r}_{n'})] , 
\label{hamgen}
\end{equation}
where $m$ is the effective carrier mass, $V(\mathbf{r})$ the Coulomb interaction, $U(\mathbf{r})$ the moiré potential and $n$ enumerates the charges. The dielectric environment and the finite thickness of the TMD layer(s) is taken into account by considering a realistically screened Coulomb interaction in real space within a generalized Keldysh approach \cite{PhysRevB.84.085406, rytova1967screened, keldysh1979coulomb}. Furthermore, the moiré potential is obtained by smoothly interpolating polarization-induced alignment shifts at high-symmetry stackings from \emph{ab initio} calculations \cite{hagel2021exciton}. Additional details on the material-specific modeling of the Coulomb repulsion and the moiré potential are provided in the Supplementary Material. 

Now, we consider the case of a charge-ordered system. The general Hamiltonian introduced in Eq. \eqref{hamgen} can then be expressed in terms of real-space Wigner lattice vectors $\mathbf{R}_n$ and deviations from the equilibrium lattice positions $\mathbf{u}_n$ by taking $\mathbf{r}_n\rightarrow \mathbf{R}_n+\mathbf{u}_n$. We are considering triangular lattices and systems with one carrier (with fixed spin polarization) per Wigner crystal cell, but note that also other geometries are possible \cite{yu2023phonons, goldoni1996stability, bonsall1977some} and could be treated in the developed framework. We assume that the deviation from the equilibrium positions is small compared to the Wigner lattice constant $a_W\sim n_e^{-1/2}$ determined by the carrier concentration $n_e$ (Fig. \ref{schematicfig}(a)) in a TMD monolayer or the moiré period set by the twist angle and/or lattice mismatch in a TMD bilayer (Fig. \ref{schematicfig}(b)). Thus, the Hamiltonian can be expanded up to the second order in the lattice deviation according to $H=H^{(0)}+H^{(2)}$, where $H^{(0)}$ is a constant contribution, $H^{(1)}$ vanishes within the harmonic approximation, and the second order contribution reads
\begin{equation}
H^{(2)}=-\frac{\hbar^2}{2m}\sum_{\mathbf{k}}\frac{\partial^2}{\partial \mathbf{u}^2_{\mathbf{k}}}+\frac{m}{2}\sum_{\alpha, \beta, \mathbf{k}}u_{\mathbf{k}, \alpha}D_{\alpha\beta}(\mathbf{k})u_{-\mathbf{k}, \beta}\ , 
\end{equation}
with the momentum $\mathbf{k}$ and the discrete Fourier transform of the lattice deviation $\mathbf{u}_{\mathbf{k}}=\frac{1}{\sqrt{N}}\sum_{n}\mathrm{e}^{-i\mathbf{k}\cdot\mathbf{R}_n}\mathbf{u}_n$ and $\mathbf{u}_{\mathbf{k}}=(u_{\mathbf{k}, x},u_{\mathbf{k}, y} )$. Furthermore, we have introduced the dynamical matrix $\mathbf{D}(\mathbf{k})$, whose elements $[\mathbf{D}(\mathbf{k})]_{\alpha\beta}\equiv D_{\alpha\beta}(\mathbf{k})$ read
\begin{equation}
D_{\alpha\beta}(\mathbf{k})=\omega_0^2\delta_{\alpha, \beta}+\frac{1}{m}\sum_{n\neq0}(1-\mathrm{e}^{i\mathbf{k}\cdot\mathbf{R}_n})\partial_{\alpha}\partial_{\beta}V(\mathbf{R}_{n}) \ .
\label{dynmat}
\end{equation}
Here, the moiré confinement frequency $\omega_0=\sqrt{\frac{8\pi^2U_0}{3ma_M^2}}$ enters, which is directly obtained via a harmonic approximation of the moiré potential, i.e. $U(\mathbf{r})\approx{\frac{m\omega_0^2}{2}r^2-U_0}$ with $U_0$ being the depth of the moiré potential  \cite{brem2024optical}. The resulting phonon dispersion $\hbar\omega_{\mathbf{k}i}$ with $i=1,2$ enumerating the phonon modes is obtained from diagonalizing the two-dimensional dynamical matrix \cite{bonsall1977some}. Note that special care has to be taken when evaluating the dynamical matrix, where due to the long-range nature of the Coulomb interaction, the summation over real-space lattice vectors in Eq. \eqref{dynmat} is slowly convergent  \cite{bonsall1977some}. Additional details on the treatment of the dynamical matrix and the diagonalization procedure are provided in the Supplementary Material.   

\begin{figure}[t!]
    \centering
    \includegraphics[width=\linewidth]{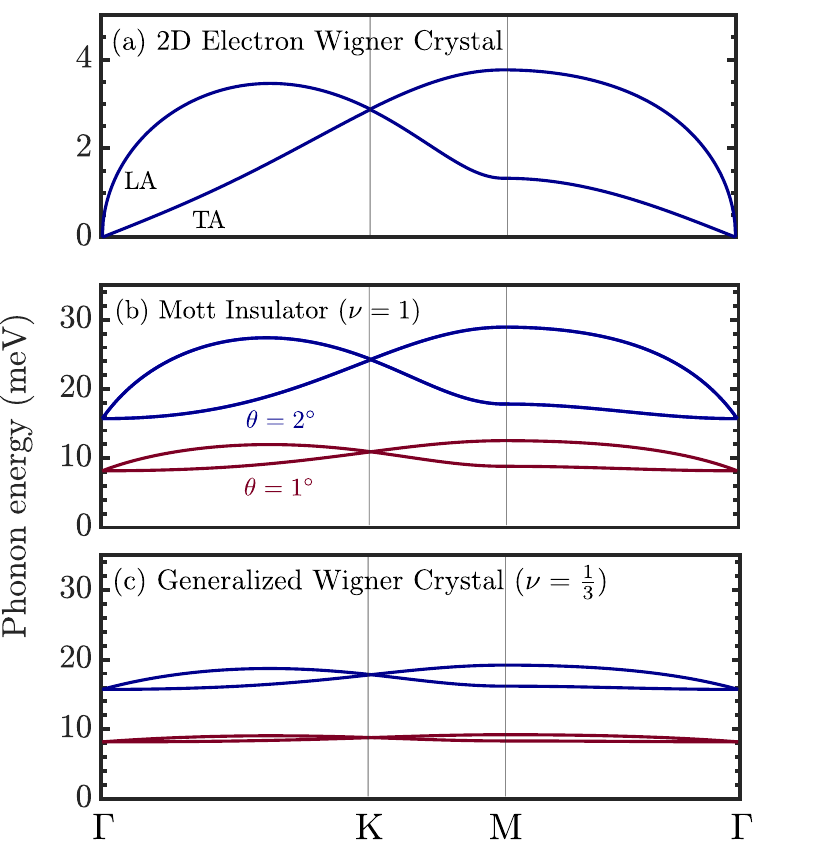}
    \caption{Phonon dispersion of  \textbf{(a)} electron Wigner crystals formed in hBN-encapsulated MoSe$_2$ monolayers with a fixed electron density of $n_e=10^{11}$ $\mathrm{cm}^{-2}$, \textbf{(b)}  Mott insulating states in twisted MoSe$_2$-WSe$_2$ heterostructures (H-type stacking, filling factor $\nu=1$), and \textbf{(c)} generalized Wigner crystals (filling factor $\nu=\frac{1}{3}$). The latter two are shown for two different twist angles.}
    \label{phondis}
\end{figure}

Figure \ref{phondis}(a) shows the phonon dispersion of a 2D electron Wigner crystal along the high-symmetry directions in reciprocal space as obtained from diagonalizing $\mathbf{D}(\mathbf{k})$ with matrix elements given in Eq. \eqref{dynmat}. Here, we consider the exemplary hBN-encapsulated MoSe$_2$ monolayer at the fixed electron density $n_e=10^{11}$ $\mathrm{cm}^{-2}$. We obtain a transverse acoustical (TA) and a longitudinal acoustical (LA) mode exhibiting a $\sim k$ and a $\sim k^{1/2}$ dependence, respectively, around the $\Gamma$ point---in agreement with previous literature on 2D Wigner crystals \cite{bonsall1977some, PhysRevA.8.2136}. As the obtained phonon energies are $<5$ meV and the density of low-energy phonon modes is large, it is anticipated that 2D electron Wigner crystals in monolayers are easily melted. In particular, just a small increase in the lattice temperature is sufficient to intensify the crystal vibrations and give rise to large fluctuations of the vibrations around their equilibrium positions. 

In contrast, considering charge-ordered states of electrons in the molybdenum layer of H-type stacked MoSe$_2$-WSe$_2$ heterostructures (Fig. \ref{phondis}(b)-(c)), which exhibit a small moiré potential depth of $U_0\sim 6$ meV (cf. Supplementary Material for details), we already find substantially larger phonon energies in the range of tens of meV. This reflects the trapping of charges in moiré pockets, 
which results in non-vanishing phonon frequencies at the $\Gamma$ point, $\omega_{\Gamma}=\omega_0$.
Since the moiré period follows $a_M\sim \frac{a_0}{\theta}$, the trapping frequency $\omega_0 \propto 1/a_M$ increases with the twist angle $\theta$.  In order to achieve the desired fillings, the carrier density needs to be tuned to the value $n_e=\nu n_\text{moir\'e}$, where $n_\text{moir\'e}=2/\sqrt{3}a_M^2$ is the density of moir\'e cells ($n_\text{moir\'e}=3.5 \cdot 10^{11}\mathrm{cm}^{-2}$ at $\theta=1^{\circ}$ and $n_\text{moir\'e}=2.85 \cdot 10^{12}\mathrm{cm}^{-2}$ at $\theta=2^{\circ}$). Here, the filling factor $\nu=1$ corresponds to the Mott insulating case (Fig. \ref{phondis}(b)) and $\nu=\frac{1}{3}$ to the generalized Wigner crystal state (Fig. \ref{phondis}(c)). 

The stronger Coulomb repulsion at higher densities (i.e. larger twist angles) results in more dispersive phonon bands and boosted elastic constants (determined by the slopes of the phonon modes near the $\Gamma$-point \cite{bonsall1977some}). In particular, the phonon energies at finite momenta can be associated with the deformation energy of the crystal which is proportional to the inter-particle forces. Consequently, the phonon dispersion becomes steeper at higher densities, where the Coulomb forces are larger. Hence, a flatter phonon dispersion is obtained at lower twist angles (lower carrier densities). Importantly, a flattening of the phonon dispersion is also seen when comparing the dispersion of a generalized Wigner crystal state ($\nu=\frac{1}{3}$) with the Mott insulating state ($\nu=1$), where the former has a lower carrier density. 
Having a microscopic access to the phonon dispersion in  TMD mono- and bilayers allows us to study the stability of charge-ordered states. Importantly, the phonon dispersion will be shown to determine the strength of thermal fluctuations of the Wigner lattice that directly enter the Lindemann melting criterion.  \\

\begin{figure*}[t!]
  \includegraphics[width=\textwidth]{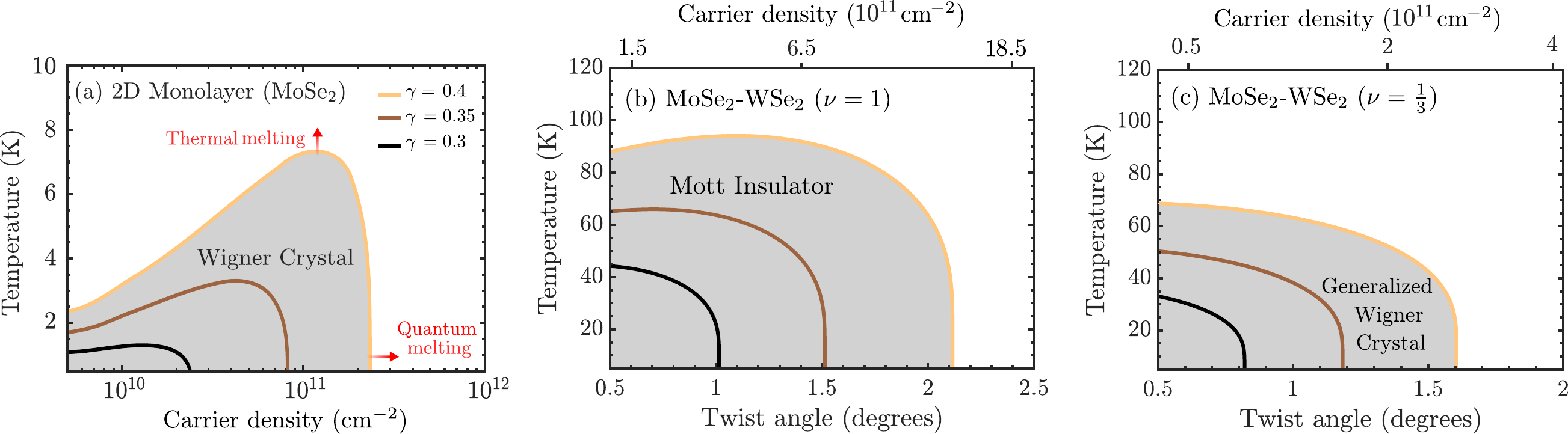}
  \caption{Phase diagrams for charge-ordered phases in TMD mono- and bilayers evaluated at fixed critical Lindemann parameters, $\gamma=0.3, 0.35, 0.4$ including the \textbf{(a)} Wigner crystal phase in hBN-encapsulated MoSe$_2$ monolayers as well as the \textbf{(b) } Mott insulating  and the \textbf{(c)} generalized Wigner crystal phase in the  hBN-encapsulated MoSe$_2$-WSe$_2$ heterostructure. 
  The grey shaded areas illustrate the maximum temperature and density range, where the corresponding material phase is stable.}
  \label{phasediagrams}
\end{figure*}

\section{Phase diagrams}
The Lindemann criterion has been widely used to study the melting of crystals and provides a qualitative understanding of the stability of charge-ordered phases across a wide range of material systems \cite{zhou2021bilayer, goldoni1996stability, smolenski2021signatures, robbins1988phase}. In general, the criterion states that melting occurs when the mean squared displacement of a particle around its equilibrium position $\langle u_n^2\rangle$ exceeds a certain fraction of the inter-particle distance $r_0$. The mean squared displacement is computed within the harmonic approximation and reads for the triangular lattice \cite{togo2015first}
\begin{equation}
\langle u^2_n\rangle= \frac{A_{\mathrm{W}} \hbar}{\pi^2m}\sum_{i}\int_{\mathrm{BZ}}\mathrm{d}^2\mathbf{k}\frac{1}{\omega_{\mathbf{k}i}}\left(N(\omega_{\mathbf{k}i},T)+\frac{1}{2}\right)
\label{msdphon}
\end{equation}
where $A_W=\frac{\sqrt{3}a_W^2}{2}$ is the Wigner crystal's unit cell area and $N(\omega_{\mathbf{k}i}, T)$ is the frequency- and temperature-dependent Bose distribution of phonons that crucially depends on the dispersion $\hbar\omega_{i\mathbf{k}}$. Note that $\langle u^2_n\rangle$ diverges with the system size for a two-dimensional electron Wigner crystal due to the long-wavelength phonons having a dispersion $\omega_k\sim k$ for the TA mode (Fig. \ref{phondis}(a)) \cite{monarkha2012two}. To overcome this issue, a modified Lindemann criterion has been introduced that compares the relative mean squared displacement between two neighbouring sites with the inter-particle separation, i.e. $\frac{\langle|\mathbf{u}_{n+1}-\mathbf{u}_n|^2\rangle}{r^2_0}=\gamma^2$ with $\gamma$ being the Lindemann parameter \cite{PhysRevResearch.2.012040, illing2017mermin, bedanov1985modified, monarkha2012two}. In this work, we use the modified Lindemann criterion to evaluate the stability of the Wigner crystal phase  in TMD monolayers, while we use the regular Lindemann criterion  for moiré structures, where the moiré confinement renders $\langle u_n^2\rangle$ finite. Remarkably, the critical Lindemann parameter $\gamma$ at which the melting of a crystal phase occurs is known to be quite insensitive to the form of both the interaction potential and the dimensionality of the considered system \cite{bedanov1985modified, robbins1988phase}, with $\gamma\gtrsim 0.3$ for 2D electron crystals \cite{bedanov1985modified}. 
With the density- and temperature-dependent mean squared displacements at hand, it remains to specify the inter-particle separation $r_0$. For the case of the 2D electron crystal, the separation $r_0=a_W$ is given by the Wigner lattice constant $a_W$ (Fig. \ref{schematicfig}(a)), and for the Mott insulating case in a heterobilayer (Fig. \ref{schematicfig}(b)), it holds that $r_0=a_M$ with the moir\'e lattice constant $a_M$. Note, that for the generalized Wigner crystal, we also set $r_0=a_M$ rather than the Wigner lattice constant, since the crystal phase is considered melted as soon as an electron jumps to a neighbouring (unoccupied) moiré pocket, which costs less energy compared to the hopping to another Wigner lattice site. 

We now evaluate the phase diagrams of charge-ordered states at fixed critical Lindemann parameters using the modified (regular) Lindemann criterion for TMD monolayers (TMD moiré bilayers). The calculated phase diagrams are presented in Fig. \ref{phasediagrams}. Here, the melting temperature and quantum melting carrier density are evaluated for a fixed range of critical Lindemann parameters $\gamma\in [0.3, 0.4]$ covering a range obtained from molecular dynamics simulations \cite{bedanov1985modified, PhysRevB.89.094112}. Note that the latter are based on  the modified Lindemann criterion. As discussed in Ref. \cite{monarkha2012two}, for high-energy phonons whose motion is uncorrelated between different nearest neighbors ($\langle u_n u_{n+1}\rangle \simeq 0$), the critical parameters based on the regular  Lindemann criterion differ by a factor $\sqrt{2}$. Therefore, for reasons of consistency, we take $\gamma'=\gamma/\sqrt{2}$  as critical Lindemann parameter for the bilayer case.

Evaluating the phase diagram for the electron Wigner crystal in hBN-encapsulated MoSe$_2$ monolayer (Fig. \ref{phasediagrams}(a)),  we obtain a melting temperature $T_m\sim 3 $ K and quantum melting occurs at $n_e\sim 8\cdot 10^{10}$ $\mathrm{cm}^{-2}$ assuming the critical Lindemann parameter $\gamma=0.35$. The low melting temperature directly reflects the high density of low-frequency phonon modes (Fig. \ref{phondis}(a)) that lead to large fluctuations in mean squared displacements (Eq. \eqref{msdphon}) upon just a small increase in lattice temperature. Generally, the obtained melting temperatures ($T_m\sim 1-8$ K) and densities ($n_e\sim 2.5\cdot10^{10}-2.3\cdot 10^{11}$ $\mathrm{cm}^{-2}$)  are in  good agreement with recent optical spectroscopy measurements on Wigner crystal states in MoSe$_2$ monolayers with  $T_m\sim 10 $ K at the carrier density $n_e=1.6\cdot 10^{11}$ $\mathrm{cm}^{-2}$ \cite{smolenski2021signatures}. Note that a perfect agreement cannot be expected due to the uncertainty in the value of the critical Lindemann parameter. In addition, the Lindemann criterion relies on the harmonic approximation, which does not include anharmonic effects which are known to become important when the crystal vibrations are large and the crystal is close to melting \cite{PhysRevB.26.5009}. Furthermore, the theory does not take into account disorder present in many experiments, which is known to pin the Wigner crystal state and enhance its stability\cite{huang2024electronic, PhysRevB.106.L121103}. 

Figure \ref{phasediagrams}(b)-(c) shows the twist- and temperature-dependent phase diagrams for the Mott insulating ($\nu=1$) state and the generalized Wigner crystal state ($\nu=\frac{1}{3}$) in twisted MoSe$_2$-WSe$_2$ heterostructures (with H-type stacking). Interestingly, we find a tenfold increase in the stability compared to the regular Wigner crystal in the monolayer case, both in terms of melting temperatures ($T_m\sim 45-90$ K for $\nu=1$ and $T_m\sim 35-70$ K for $\nu=\frac{1}{3}$) and the carrier densities needed for quantum melting. This reflects the substantial impact of moiré trapping in the considered heterobilayer and the overall predominant role of the moiré potential compared to the Coulomb repulsion for the phonon dispersion (Fig. \ref{phondis}(b)-\ref{phondis}(c)), where the repulsion modulates the dispersion only at larger momenta. In particular, the moiré confinement energy ($\sim \omega_0$ in Eq. \eqref{dynmat}) that is acquired by the carriers in heterobilayers requires large thermal energies $k_B T$ for the activation of vibrations as the phonon density of states of these crystals is shifted to higher energies. 
The reduced density of electrons at the filling $\nu=\frac{1}{3}$ and the resulting weakened Coulomb repulsion for the generalized Wigner crystal state leads to lower melting temperatures ($T_m\sim 50 $ K) compared to the Mott insulating phase ($T_m\sim 65 $ K), corresponding to a reduction of  approximately 25 $\%$. The enhanced stability of charge-ordered states in TMD heterobilayers  has also been experimentally measured and melting temperatures $T_m\gtrsim100$ K and $T_m\gtrsim 50$ K have been obtained for Mott insulating states and generalized Wigner crystal states \cite{xu2020correlated}---in line with our theoretical predictions. While previous studies on charge-ordered states have been reported for WS$_2$-WSe$_2$ heterobilayers \cite{xu2020correlated}, we expect our microscopic insights to be valid also in other TMD heterostructures. \\

\section{Tunability of charge-ordered states}
Temperature and carrier density represent the most important tuning knobs for controlling charge-ordered states of matter in 2D materials. In addition, at fixed filling fractions in TMD bilayers, twist-angle engineering can be used to generate correlated states across a wide range of carrier densities. Dielectric engineering of the Coulomb interaction is another important and experimentally accessible tuning strategy, which can significantly impact the stability of these  material phases \cite{raja2017coulomb, steinleitner2018dielectric}. Furthermore, the stacking configuration in heterobilayers crucially determines the  moiré potential depth and thus also plays an important role for the stability of generalized Wigner crystals and Mott insulators \cite{yuan2020twist, PhysRevB.97.035306}. 

\begin{figure}[t!]
    \centering
    \includegraphics[width=\linewidth]{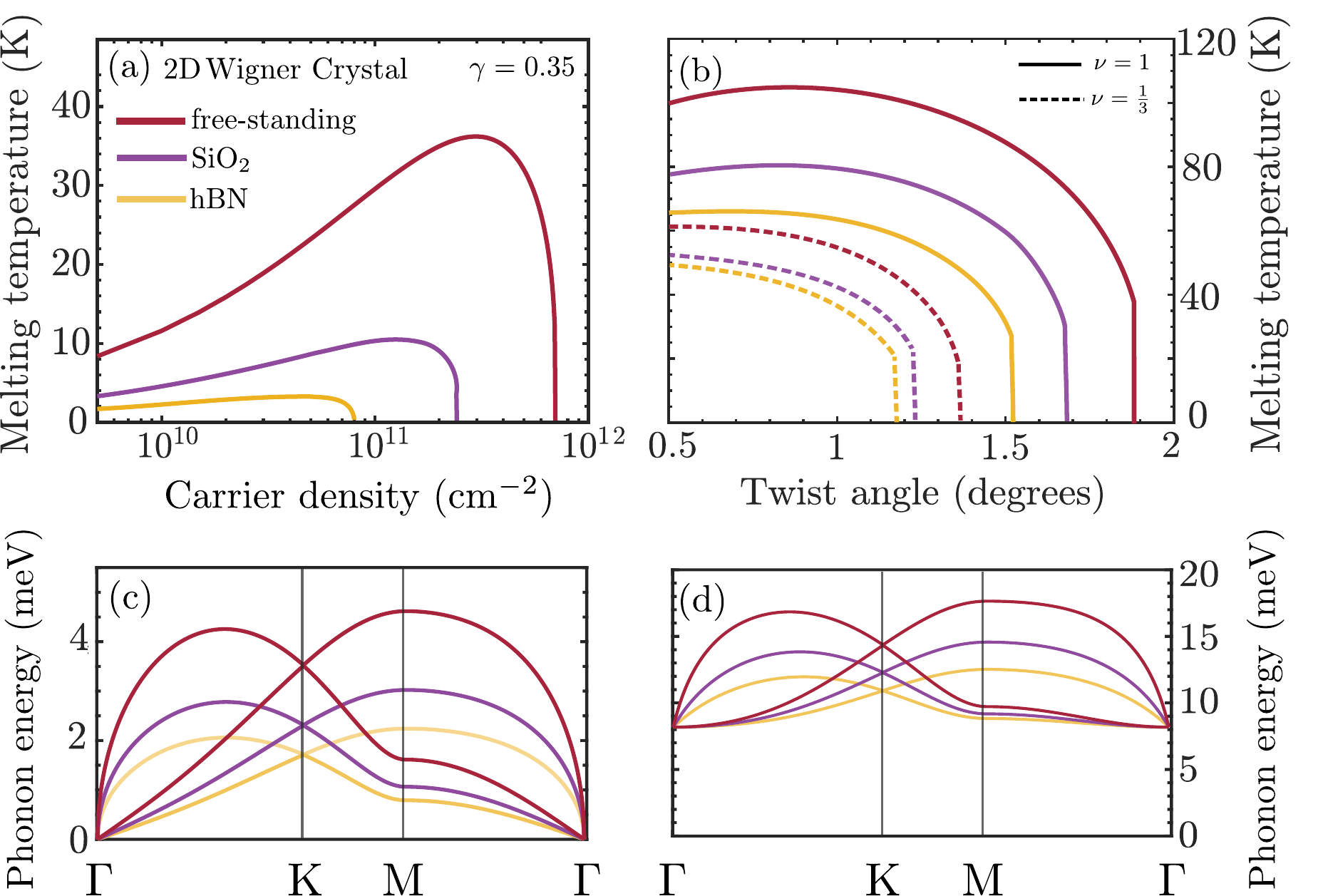}
    \caption{Dielectric engineering of charge-ordered phases. \textbf{(a)} Melting temperature of the electronic Wigner crystal phase in MoSe$_2$ monolayers as a function of electron density for different dielectric substrates with a fixed Lindemann parameter of $\gamma=0.35$. \textbf{(b)} Melting temperature of the Mott insulating phase (solid lines) and the generalized Wigner crystal (dashed lines) in hBN-encapsulated MoSe$_2$-WSe$_2$ heterostructure as a function of the twist angle.  Phonon dispersion of the \textbf{(c)} Wigner crystal phase at the electron density $n_e=5\cdot 10^{10}$ $\mathrm{cm}^{-2}$ and \textbf{(d)} the Mott insulating state  at the twist angle of $\theta=1^{\circ}$. }
    \label{screeningdep}
\end{figure}

Figure \ref{screeningdep} displays the melting temperature as a function of the carrier density for the charge-ordered phases in MoSe$_2$ monolayers (Fig. \ref{screeningdep}(a), Lindemann parameter $\gamma=0.35$) and in twisted MoSe$_2$-WSe$_2$ heterobilayers (Fig. \ref{screeningdep}(b), $\gamma'=0.35/\sqrt{2}\approx 0.25$) assuming three different choices of dielectric environments ranging from free-standing (dielectric constant $\epsilon_s=1$, red lines) to the SiO$_2$ substrate ($\epsilon_s\approx{2.45}$, purple lines) and hBN-encapsulation ($\epsilon_s\approx{4.5}$, yellow lines). Remarkably, the melting temperature of a monolayer Wigner crystal increases by more than one order of magnitude (from $T_m\sim 3$ K to $T_m\sim 40 $ K) when considering free-standing samples compared to hBN-encapsulated TMD monolayers. This is a consequence of the enhanced Coulomb repulsion between the charges with reduced dielectric screening, which causes the electrons to maintain the crystal configuration also at higher densities. A reduction of the dielectric constant also results in larger phonon energies (Fig. \ref{screeningdep}(c)), increased sound velocity of the TA mode and a stiffening of the phonon spectrum \cite{PhysRevB.30.159}. As a consequence, mean squared deviations in phonon amplitudes, i.e. harmonic oscillations of the charges, are decreased and the stability of the Wigner crystal phase in monolayers is enhanced. 

For Mott insulators (solid lines in Fig. \ref{screeningdep}(b)) and generalized Wigner crystals (dashed lines in Fig. \ref{screeningdep}(b)), the melting temperature is less sensitive to changes in the dielectric environment. 
This is due to the fact that the ``phonon gap'' induced by the moiré potential is independent of dielectric environment (Fig. \ref{screeningdep}(d)) as it stems from the moiré confinement. 
In particular, we find that melting temperatures are reduced by approximately 40 \% when going from free-standing to hBN-encapsulated samples. 
The Coulomb interaction only influences the dispersion at finite momenta, where a less screened Coulomb interaction is seen to result in more dispersive phonon energies and a smaller phonon density of states. Consequently, smaller spatial fluctuations of  charges around their equilibrium sites are obtained, overall leading to more stable Mott insulators and Wigner crystals for weakly screened TMD materials.

\begin{figure}[t!]
    \centering
    \includegraphics[width=\linewidth]{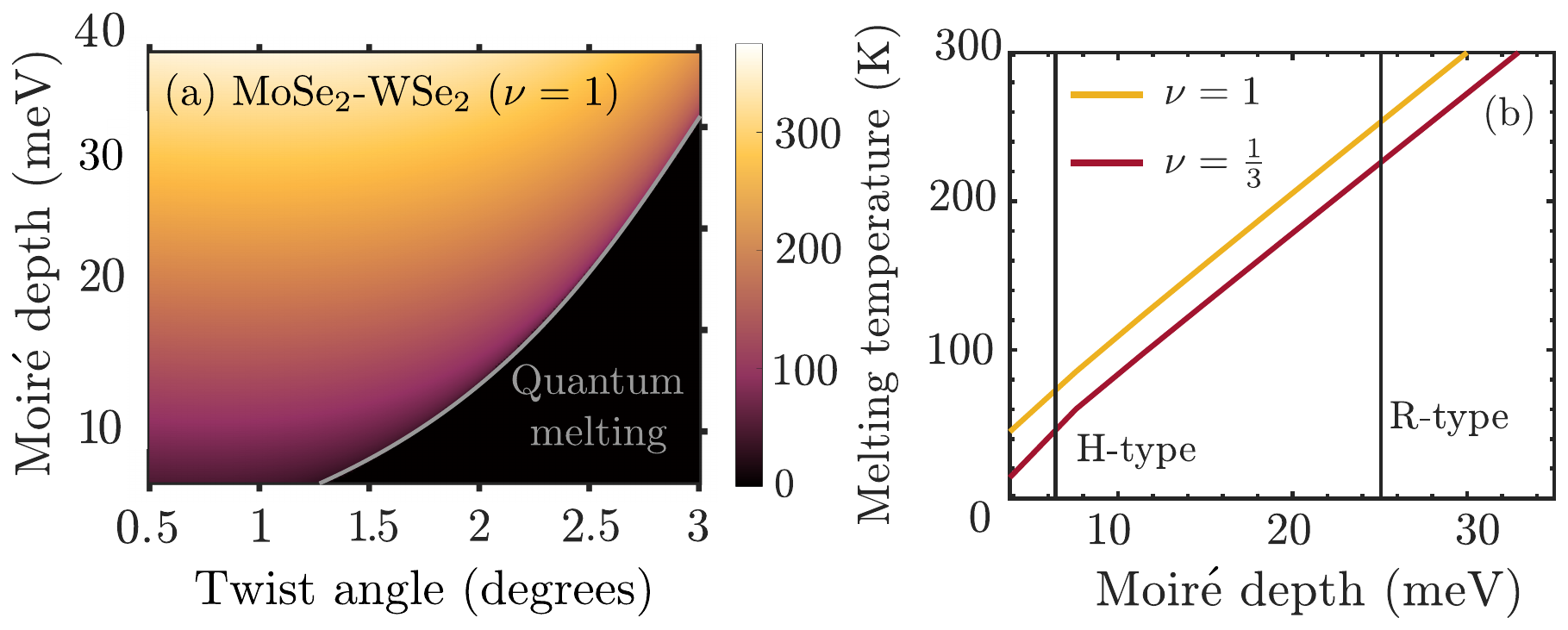}
    \caption{Impact of moiré trapping on melting temperatures for charge-ordered phases in twisted MoSe$_2$-WSe$_2$ heterostructures. 
    \textbf{(a)} Melting temperature as a function of the twist angle and moiré depth for the Mott insulating phase. \textbf{(b) } Cut at the fixed twist angle $\theta=1^{\circ}$ (including now also the generalized Wigner crystal phase). The moiré potential depths of H-type and R-type stacking are indicated by vertical black lines.}
    \label{figmoireconf}
\end{figure}

Finally, we investigate the role of stacking configuration on the stability of charge-ordered phases in TMD heterobilayers. It has been shown in first-principle studies that the moiré potential  is substantially larger at R-type (parallel) stacking than at H-type (anti-parallel) stacking with potential depths ranging from just a few meVs for H-type potentials \cite{PhysRevB.97.035306} up to several tens of meVs for R-type stacking \cite{yuan2020twist, PhysRevLett.121.026402}. Hence, we expect the stability of correlated material phases in heterobilayers to be strongly dependent on the stacking configuration. To study this effect, we treat in the following the moiré potential depth as a parameter $U_0\in [5, 40]$ meV and recalculate the melting temperature for the Mott insulating state in the twisted MoSe$_2$-WSe$_2$ heterobilayer at the fixed critical Lindemann parameter $\gamma'=0.35/\sqrt{2}\approx 0.25$, but now as a function of twist angle and moiré potential depth (Fig. \ref{figmoireconf}(a)). We find that, by tuning the potential depth by an order of magnitude, melting temperatures of $T_m \approx 300$ K can be obtained for the considered range of twist angles. Furthermore, the regime of quantum melting (black-shaded region), is pushed towards higher twist angles (i.e. smaller moiré periods) when the moiré potential is deeper. This can be directly traced back to  the  moiré confinement frequency $\omega^2_0\sim \frac{U_0}{a^2_M}$ (Eq. \eqref{dynmat}), which predominantly determines the phonon dispersion in bilayers (Fig. \ref{phondis}(b)). 
From the latter expression we obtain the approximate relation $\gamma^2\sim \frac{1}{\sqrt{U_0 a_M^2}}$ at vanishing temperatures, consequently implying that charge-ordered states which experience deeper moiré potentials (large $U_0$) can sustain larger carrier densities (small $a_M$). 

Interestingly, at finite temperatures, we reveal a linear relationship between melting temperature and moiré potential depth both for Mott insulators (yellow lines) and  generalized Wigner crystals (red lines), cf. Fig. \ref{figmoireconf}(b). 
This dependence can be qualitatively explained by assuming a flat phonon dispersion (Fig. \ref{phondis}(b)), i.e. $\omega_{\mathbf{k}i}\approx{\omega_{0}}\sim \sqrt{U_0}$. In this case, the mean squared displacement of lattice deviations (Eq. \eqref{msdphon}) can be analytically evaluated and a closed expression for the melting temperature is found according to $T_m=\hbar\omega_{0}/(k_B\mathrm{log}(1+\kappa/(\gamma^2\omega_{0}-\kappa/2)))\approx{\hbar\omega^2_{0}\gamma^2/(k_B \kappa)}\sim U_0$, where $\kappa$ is a constant. Note that so far we have focused on  MoSe$_2$-WSe$_2$ at H-type stacking (indicated with a vertical black line in Fig. \ref{figmoireconf}(b)). 
For R-type stacked MoSe$_2$-WSe$_2$ we obtain a substantially larger moiré potential depth for carriers ($U_0\approx{25}$ meV), which consequently results in an enhanced stability of charge-ordered states. The statement that the R-type twisted MoSe$_2$-WSe$_2$ heterostructure hosts more stable Mott insulating and generalized Wigner crystal states is expected to hold also for other TMD heterostructures. Note that our calculations do not take into account lattice relaxation effects due to atomic reconstruction, which can modify the moiré pattern at low twist angles ($\theta\lesssim 1^{\circ}$)  \cite{weston2020atomic, zhao2023excitons, PhysRevMaterials.8.034001}. Atomic reconstruction deforms moiré supercells into large periodic domains which results in significant strain profiles and deeper potentials \cite{enaldiev2022self}. Therefore, we expect the reconstruction to  further enhance the stability of charge-ordered phases. \\

\section{Conclusion}
In this work, we have investigated the stability of charge-ordered phases in TMD monolayers and twisted heterostructures using the Lindemann criterion with the interaction potential obtained from a microscopic theory. Our findings indicate that the moiré potential present in heterostructures makes it possible to engineer charge-ordered states even at room temperature depending on the twist angle and stacking configuration. Furthermore, we show that the phonon excitations of generalized Wigner crystal states shed light on the interplay between Coulomb repulsion and moiré pinning. Here,  the Coulomb interaction provides a stiffness to the Wigner lattice, while the moiré potential leads to harmonic traps of electrons. In contrast, the regular Wigner crystals in TMD monolayers are more fragile. However,  their melting temperatures can be significantly tuned by engineering the dielectric environment, where a tenfold increase in the melting temperatures is predicted for the weakly screened free-standing TMDs compared to the hBN-encapsulated ones. Overall, the obtained microscopic insights can guide future experiments toward the realization of more stable charge-ordered phases.  \\

\section*{Acknowledgments}
We thank Joakim Hagel, Fredrik Eriksson and Petter Rosander (Chalmers University of Technology) for useful discussions. This project has received funding from  Deutsche Forschungsgemeinschaft via CRC 1083 (Project No. B09) and the regular project 542873285.

\end{document}